# Hole - Nuclear Spin Interaction in Quantum Dots

B. Eble[1], C. Testelin[1], P. Desfonds[1], F. Bernardot[1], A. Balocchi[2], T. Amand[2], A. Miard[3], A. Lemaître[3], X. Marie[2] and M. Chamarro[1]

[1] *Institut des NanoSciences de Paris, Université P. et M. Curie, CNRS-UMR 7588, 140 rue de Lourmel, F-75015 Paris, France*

[2] *Université de Toulouse; INSA, UPS, CNRS ; LPCNO, 135 avenue de Rangueil, F-31077 Toulouse, France*

[3] *Laboratoire de Photonique et Nanostructures, CNRS, Route de Nozay, F-91460 Marcoussis, France*

*We have measured the carrier spin dynamics in p-doped InAs/GaAs quantum dots by pump-probe photo-induced circular dichroism and time-resolved photoluminescence experiments. We show that the hole spin dephasing is controlled by the hyperfine interaction between hole and nuclear spins. In the absence of external magnetic field, we find a characteristic hole spin dephasing time of 15 ns, in close agreement with our calculations based on dipole-dipole coupling between the hole and the quantum dot nuclei. Finally we demonstrate that a small external magnetic field, typically 10 mT instead of 200 mT for the case of electrons, quenches the hyperfine hole spin dephasing.*

An individual spin in a solid confined to a nanometer scale represents an ultimate quantum memory and is thus a potential candidate for the realization of spintronic and quantum information processing devices [1,2,3]. Compared to bulk materials, the strong spatial confinement of carriers in semiconductor quantum dots (QDs) efficiently quenches the main spin relaxation mechanisms such as the D'Yakonov-Perel one [4] while enhancing mechanisms based on carrier exchange and hyperfine interaction. The hyperfine interaction of an electron with the nuclei, based on the Fermi contact term, has recently been identified as the primary obstacle to quantum computation applications as it yields efficient electron spin dephasing [5,6,7,8]. For a hole, this Fermi contact coupling, is expected to be much weaker because of the p-symmetry of the valence band states [4, 9]. Indeed pioneer experimental work performed in silicon, fifty years ago, has shown that the *nuclear* spin relaxation time due to the interaction with free carriers was about one order of magnitude smaller in p-doped than in n-doped bulk material [10].

However the corresponding *hole* spin relaxation or dephasing induced by nuclear spins were not evidenced before in semiconductors because the hole spin dynamics in bulk or quantum well structures is governed by the very rapid spin relaxation mechanisms induced by the strong heavy- and light-hole mixing in the valence bands [4,11,12,13]. These spin-orbit related effects are strongly inhibited in QDs due to their fully quantized electronic structure. Recently, a lower limit of tens of nanoseconds in zero external magnetic field was estimated for hole spin relaxation times in CdSe or InAs QDs, but these measurements were limited by the radiative recombination of the photo-created complexes [14,15]. In the presence of high external magnetic fields (1-12 T), which screen the effects related to nuclear spins, it has been demonstrated that a spin-orbit-mediated single-phonon scattering between Zeeman levels governs the hole spin relaxation processes [16,17,18]. This mechanism leads to a decrease of the hole spin relaxation time when increasing the magnetic field.

In this letter, we measure the hole spin dynamics by time-resolved optical orientation experiments in an ensemble of p-doped InAs/GaAs QDs. We demonstrate that, in the absence of applied magnetic field, the hole spin dephasing is controlled by the hyperfine hole-nuclear spin interaction. We show that this effect relies on the dipole-dipole coupling combined with the mixing of heavy-and light-hole states in QDs. Finally we evidence that a small magnetic field of the order of 10 mT, instead of 200mT for the case of electrons, quenches the hole spin relaxation.

The QDs structures were grown by molecular beam epitaxy on a (001) GaAs substrate. The sample presented here consists of 30 planes of self-assembled InAs QDs, separated by 38-nm thick GaAs spacer layers. The QD surface density is about $10^{10}$ cm$^{-2}$. The structure was p-modulation doped with a Carbon delta-doping layer (nominal density ~$2\times10^{11}$ cm$^{-2}$) located below each QD layer.

To probe the resident hole spin polarization, we measured the photo-induced circular dichroism (PCD) in the QD sample. A picosecond Ti:sapphire laser is split into pump and probe beams (the repetition frequency is 76 MHz). The pump beam polarization is σ+/σ- modulated at 42 kHz with a photo-elastic modulator; the probe beam is linearly polarized. After transmission through the sample, the probe beam is decomposed into its two circular components, and the difference in their intensities is measured with a balanced optical bridge. To improve the signal-to-noise ratio, a double lock-in amplifier analysis of the signal is performed, the pump and probe beams being modulated with a mechanical chopper at two different frequencies. In the same sample, the electron spin dynamics has been measured by time-resolved photoluminescence (PL) experiments; 1.5 ps pulses generated by a Ti-Sapphire laser at a repetition frequency of 82 MHz were used as excitation light, and the PL signal was recorded by using a S1 photocathode streak camera with an overall time resolution of 20 ps.

In the PCD experiments, a circularly polarized pump beam propagating along the growth axis z is tuned to the energy of the lowest-allowed optical transition of InAs QDs, containing a single resident hole (E = 1.35 eV). This excitation creates a complex of three particles, called the positive trion ($X^+$), in its ground state. This transient complex consists of two holes with opposite spins forming a singlet (the photo-generated one and the resident one due to doping), and of a photo-generated electron with its spin pointing down or up depending on the σ⁺ or σ⁻ circularly polarized excitation, respectively. For simplicity, we consider, for the optical orientation process, pure heavy-hole states [19]. Thus, the pump beam helicity selectively generates spin-polarized electrons. During the lifetime of these photo-generated electrons, the electron-nuclear hyperfine interaction leads to an efficient coherent coupling of the two electron spin states [5,7]. The spontaneous decay of the trion state by emission of a polarized photon leads to efficient hole spin cooling, as evidenced recently in cw single QD experiments [20] (see Fig. 1a). This process allows us to spin-polarize the resident holes in the QDs with a pulsed resonant excitation.

Figure 1b shows the temporal behaviour of the low-temperature PCD signal obtained when the pump and probe beams are tuned to the trion transition of the p-doped QDs. The PCD signal has two contributions: (i) the population difference $\rho_{+3/2} - \rho_{-3/2}$ of the spin-polarized heavy-hole ground states $J_z^h = \pm 3/2$, and (ii) the population difference $\rho_{+1/2} - \rho_{-1/2}$ of the spin-polarized trion states $J_z^{X^+} = \pm 1/2$. Thus the PCD signal simply writes: $PCD \propto (\rho_{+3/2} - \rho_{-3/2}) - (\rho_{+1/2} - \rho_{-1/2})$. In Fig. 1 b, we observe a nonzero PCD signal at negative pump-probe delays, indicating that the spin polarization is not fully relaxed within the $T_L = 13$ ns repetition period of the laser pulses. This long-living component of the PCD signal is unambiguously associated to the net spin polarization of the resident holes, the only species present in the sample after the radiative recombination of trions ($T_R \approx 800\,ps$, see Fig. 3).

To obtain a direct evidence of the hole-nuclear spin interaction, we have applied an external magnetic field $\vec{B}$ parallel to the growth axis $z$ of the sample. As for an electron [5,6], the hyperfine interaction of a hole in a QD with the surrounding nuclei can be described by a frozen effective nuclear field $\vec{B}_N^h$ acting on the hole spin. This frozen fluctuation approach is justified since the correlation time of the field $\vec{B}_N^h$ (~100 μs) is several orders of magnitude longer than the typical hole spin dephasing time, as will be found later. The only difference is the physical origin of this hyperfine interaction. For an electron, the hyperfine interaction has a dominant Fermi contact character because of the s-symmetry of the wavefunction. However, a hole in the valence band possesses a p-symmetry (the admixture with s-symmetry conduction wavefunction being negligible [21]), the wavefunction at the nuclei vanishes and the contact interaction is thus very weak. This is the reason why the hole-nuclear spin interaction has been neglected so far. Nonetheless, the nuclei can also interact with carriers through dipole-dipole coupling [22] as we will show and calculate later. In the absence of applied magnetic field, each resident hole spin precesses coherently around the effective nuclear magnetic field $\vec{B}_N^h$, as a consequence of the dipole-dipole hyperfine interaction (see Fig. 2a). The average hole spin polarization in the QDs ensemble thus decays with time because of the random distribution of the local nuclear effective fields. In an external magnetic field, the hole spin dephasing induced by the hyperfine interaction can be strongly reduced if the amplitude of the external field $\vec{B}$ is larger than the dispersion $\Delta_{II}$ of the in-plane fluctuations of the nuclear hyperfine field $\vec{B}_N^h$ (i.e. the Zeeman interaction of the hole spin with $\vec{B}$ is stronger than the interaction with nuclei).

Figure 2b shows the PCD signal obtained for several values of the applied magnetic field $\vec{B}$. The mean pump power density is ≈ 60 W.cm$^{-2}$, close to the previously estimated power density of Rabi π–pulses in n-doped InAs QD [23,24]. We clearly see in figure 2b that the amplitude of the PCD signal for t<0 increases significantly with the increase of B. Figure 2 c shows the PCD signal measured at delay t=-130 ps, as a function of the applied magnetic field, with a Full Width at Half Maximum (FWHM) 2Δ= 5 ± 0.4 mT. The striking feature is that a small external field has a dramatic impact on the resident hole spin polarization. The observed rise of the measured PCD(t=-130 ps ≈13ns) signal when B increases reflects a significant increase of the hole spin polarization due to a longer hole spin dephasing time. We emphasize that the external magnetic field $B_z$ remains very small (*i.e.* of the order of few mT; the Zeeman splitting of the electron or the hole in this field is about 3 orders of magnitude smaller than $k_B T$ at T= 2 K). Besides, the rapid $\sigma+/\sigma-$ modulation (at 42 kHz) of the pump beam in the experiment prevents any dynamical polarization of the nuclear spins [4,25]. Therefore, the increase of the average hole spin polarization observed in Fig. 2c is interpreted as a consequence of the suppression of the hole spin dephasing induced by the interaction with nuclear spins: the hole spin relaxation time becomes much longer than the laser repetition period ($T_L$~13 ns). Hence, the PCD signal at negative delays corresponds to an average hole spin polarization which results from the equilibrium between the laser repetition period and the relevant hole spin relaxation time $T_1^h$ (no longer due to the hyperfine interaction with nuclei if $B_z$ is larger than 10 mT).

To confirm this interpretation, we have calculated the hole – nuclear spin hyperfine interaction through dipole-dipole coupling and we show below that it explains the magnetic field dependence of the PCD signal at 13 ns displayed in Fig 2c. In a QD, confinement and biaxial strains lift the degeneracy between the heavy-holes (hh) states, $\varphi_{\pm 3/2} = |J=3/2, J_z = \pm 3/2\rangle$, and the light-holes (lh) states, $\varphi_{\pm 1/2} = |J=3/2, J_z = \pm 1/2\rangle$. In the case of pure hh states, the dipole-dipole hyperfine hamiltonian does not couple $\varphi_{+3/2}$ and $\varphi_{-3/2}$. Nonetheless, there are clear experimental evidences of heavy and light hole mixing of the hole wavefunction of self-assembled QDs induced by in-plane anisotropy and strain [26,27,28]. The modified hh states can be written as [28,29] $\tilde{\varphi}_{\pm 3/2} = \frac{1}{\sqrt{1+|\beta|^2}}(\varphi_{\pm 3/2} \mp \beta\, \varphi_{\mp 1/2})$, where $\beta = |\beta| e^{i\delta}$ is a coefficient taking into account the strain contribution. Then, the general dipole-dipole Hamiltonian [22] for a given hole can be written as follows in the basis $(\tilde{\varphi}_{+3/2}, \tilde{\varphi}_{-3/2})$:

$$H_{dd} = \Omega \sum_j \frac{C_j}{1+|\beta|^2} |\Psi(\vec{R}_j)|^2 \left[\alpha\left(\tilde{I}_x^j S_x + \tilde{I}_y^j S_y\right) + \tilde{I}_z^j S_z\right]. \quad (1)$$

The summation is over all the lattice nuclei with magnetic momentum $\mu_I^j$, spin $\vec{I}^j$ and position $\vec{R}_j$. Here $\Omega$ is the volume of a primitive unit cell (containing two nuclei for InAs or GaAs compounds). $\vec{S}$ is a pseudo-spin 1/2, whose eigen values $S_z = \pm 1/2$ are associated to the heavy-hole $\tilde{\varphi}_{\pm 3/2}$ states. $\tilde{I}_{x,y,z}$ denotes : $\tilde{I}_x^j = I_x^j \cos\delta + I_y^j \sin\delta$, $\tilde{I}_y^j = -I_x^j \sin\delta + I_y^j \cos\delta$ and $\tilde{I}_z^j = I_z^j$; $\Psi(\vec{R}_j)$ is the hole envelope wave function at the j$^{th}$ nucleus, and $\alpha = \frac{2|\beta|}{\sqrt{3}}$ is the anisotropy constant. Typical values of $|\beta| = 0.2\text{-}0.7$ have been observed in InAs, CdSe or CdTe QDs [26,27,28]. The dipole-dipole hyperfine constants $C_j$ can be obtained by using the dipole-dipole matrix elements derived in ref.[9], with symmetry rules applied to the hole Bloch function. As in ref.[9], the main contribution to the dipole-dipole interaction is due to the short-range contribution of the dipole-dipole coupling. This leads to $C_j = \frac{16}{5}\frac{\mu_B \mu_I^j}{I^j}\left\langle\frac{1}{r^3}\right\rangle_j$, where $\left\langle\frac{1}{r^3}\right\rangle_j$ denotes the average of $\frac{1}{r^3}$ over a unit cell ($\vec{r}$ being centered on nucleus $j$) and $\mu_B$ is the Bohr magneton. From ref.[30], one gets $\left\langle\frac{1}{r^3}\right\rangle_{As}$ = (4.8 ± 0.3) x 10$^{25}$ cm$^{-3}$ and $\left\langle\frac{1}{r^3}\right\rangle_{In}$ = (3.4 ± 0.5) x10$^{25}$ cm$^{-3}$. This leads to $C_{As}$ = 4.4 ± 0.3 µeV and $C_{In}$ = 4.0 ± 0.5 µeV.

The effective nuclear field $\vec{B}_N^h$ experienced by a hole in a QD, through dipole-dipole interaction with the nuclei, is then written:

$$\vec{B}_N^h = \frac{\Omega}{g_h \mu_B} \sum_j \frac{C_j}{1+|\beta|^2} |\Psi(\vec{R}_j)|^2 \left[ \alpha \left( \tilde{I}_x^j \vec{e}_x + \tilde{I}_y^j \vec{e}_y \right) + \tilde{I}_z^j \vec{e}_z \right], \quad (2)$$

where $g_h$ is the hole Landé factor, and $\vec{e}_x, \vec{e}_y, \vec{e}_z$ are the unit vectors along the x,y, z axes. The magnitude and direction of this field are randomly distributed from QD to another QD, and the randomness is described by an anisotropic gaussian probability distribution of $\vec{B}_N^h$ with $\Delta_{//}$ and $\Delta_\perp$ the quadratic averages of the in-plane and perpendicular -to- the plane components. Assuming a simple model with a constant wavefunction inside the QD and equal to zero outside, $\Delta_{//}$ and $\Delta_\perp$ write :

$$g_h \mu_B \Delta_{//} = \alpha \ g_h \mu_B \Delta_\perp = \frac{\alpha}{1+|\beta|^2} \sqrt{\frac{4\sum_j I^j(I^j+1)(C_j)^2}{3N_L}} = \frac{\hbar}{T_\Delta^h}, \quad (3)$$

where $T_\Delta^h$ is the ensemble dephasing time, arising from the random hole precession directions and frequencies in the randomly distributed frozen nuclear field. $N_L$ is the number of nuclei inside a QD, and the summation runs over the nuclei j inside a unit cell.

To obtain information about the electron-nuclear coupling strength in the same sample, we measured the photo-created electron spin dynamics in time-resolved photoluminescence (PL) experiments. Figure 3 displays the dynamics of the circular polarization $P_c = \frac{I^+ - I^-}{I^+ + I^-}$ detected at the peak of the QD ground state PL (1.36 eV) after a circularly-polarized $\sigma^+$ picosecond excitation ($I^+$ and $I^-$ are the luminescence intensities, co-polarized and counter-polarized with the circularly polarized laser pulse). This curve allows us to measure the time evolution of the average $X^+$ spin, i.e. the photo-generated electron spin since the two holes in the trion form a spin singlet. As demonstrated before [7,8], the observed decay of the PL circular polarization directly reflects the spin dephasing of the photo-generated electrons due to their hyperfine interaction with nuclei. This interaction can be described as an effective nuclear field $\vec{B}_N^e$ acting on a localized electron spin, which is frozen during the lifetime of the photo-generated electron [5]. The initial decay time of the average electron spin polarization, $T_\Delta^e$, which characterizes the fluctuations of the hyperfine interaction can be estimated from Fig.3 to be $T_\Delta^e \sim 500$ ps, which is comparable to previous results in similar systems. By using the contact hyperfine constants $A_{As} = 46$ μeV and $A_{In} = 56$ μeV from ref. [31], and the theoretical expression for $T_\Delta^e$, we obtain a value of $N_L \approx 6 \times 10^4$ which is in good agreement with the average size of the QDs. From the measured value of $T_\Delta^e$, we can get the $\Delta_B^e$ parameter, related to the fluctuations of the nuclear field acting on the localized electron spins; we find $\Delta_B^e = \frac{\hbar}{g_e \mu_B T_\Delta^e} = 57$ mT with an average electron Landé factor $|g_e| = 0.4$, measured in the same sample from the frequency of spin quantum beats of the PL signal in a tilted magnetic field geometry (not shown) [32,33].

Finally, taking the previously estimated values of $C_j$ and the value of $N_L$ deduced from the $T_\Delta^e$ measurement, we are now able to calculate the hole spin dephasing time $T_\Delta^h$ from Eq. (3); for a typical value of the anisotropy constant observed experimentally [26,27,28] $|\beta| = 0.6$, we find $T_\Delta^h = 13$ ns. That leads to a quadratic average characterized by $2\Delta_{//} = 1,2$ mT (assuming $g_h \approx 1.5$). We note that this value is roughly of the same order of magnitude but smaller than the FWHM of the experimental curve in Fig. 2 c. In fact the width of the experimental curve does not correspond directly to $2\Delta_{//}$ : the width increases when the observation time becomes comparable or shorter than the spin dephasing time, as already calculated by Merkulov *et al.* for the magnetic field dependence of the electron spin polarization [4] . To illustrate this point, we have used the spin dynamics model developed by these authors for an assembly of spin-polarized electrons, and we have adapted it to describe the spin dynamics of an assembly of holes initially polarized along z direction and coupled to random nuclear fields with an anisotropic distribution. We have calculated the magnetic field dependence of the ensemble spin polarization at t = 13 ns, for typical values of $|\beta|$ and $g_h$ (see the inset of Fig. 2c). Despite its simplicity (the effect of the periodic laser excitations has not been considered here), this calculation reproduces quite satisfactorily the measured hole spin behaviour under longitudinal magnetic field. We note that the calculated FWHM is about $2\Delta_{//}$(calc)~4 mT, close to the measured value in figure 2c.

In conclusion, we have demonstrated that the hole spin dynamics in InAs semiconductor QDs is governed by the dipole-dipole interaction with randomly oriented nuclear spins due to the heavy- and light-hole mixing of the QD ground state. The effect of this hyperfine interaction on the hole spin relaxation time can be efficiently suppressed by an external magnetic field provided that it is larger than the fluctuations of the effective nuclear field acting on the hole spin. We have shown that this field is of the order of a few mT, which is one order of magnitude smaller than the magnetic field required to screen the interaction of an electron with the nuclear spins in the same dots. This nuclear induced hole spin dephasing, which has not been measured before in semiconductors, must be taken into account in future nanoscopic hole-spin-based quantum devices. Finally, studies to control and minimize the heavy-light hole mixing would be interesting in order to reduce or even cancel the hyperfine interaction of holes in QDs.

We acknowledge O. Krebs and B. Urbaszek for fruitful discussions. One of us (B.E.) thanks the C'Nano-IdF for its financial support.

Figure captions

Figure 1 :

a) Schematic representation of the hole spin polarization mechanism: a σ⁺ circularly-polarized pulse (solid line) creates a positive trion $|\Downarrow\Uparrow\downarrow\rangle$, whose electron spin can precess in the effective nuclear field $\vec{B}_N^e$ induced by the Fermi contact hyperfine coupling with the dot nuclei. This process generates a $|\Downarrow\Uparrow\uparrow\rangle$ trion. Then, through spontaneous recombination (wavy line) the $|\Downarrow\Uparrow\uparrow\rangle$ trion is shelved in the $|\Uparrow\rangle$ state, leading to hole polarization. The hole spin can then relax due to the effective nuclear field $\vec{B}_N^h$ induced by dipolar hyperfine coupling.

b) PCD signal as a function of the pump-probe delay, at zero external magnetic field. The pump and probe energies are tuned to the lowest excited state of the p-doped InAs/GaAs QDs (1.35 eV). T=2K. Inset: schematic representation of the PCD measurement.

Figure 2

a) Schematic representation of the dipole-dipole hyperfine interaction of a single resident hole with the surrounding nuclei localized within the hole envelope wavefunction. In the mean-field approximation, the effect of the nuclei on the QD hole spin is described by the hole spin precession in an effective magnetic field $\vec{B}_N^h$ with random direction and modulus from dot to dot. As a consequence of the random nuclear field, the ensemble hole spin polarization relaxes.

b) PCD signal as a function of the pump-probe delay, for different values of the external magnetic field $\vec{B}$ applied along the sample growth axis (z-direction). The pump and probe beams beams are tuned to 1.35 eV. The zero-signal level is the same for all displayed curves. T = 2K

c) PCD amplitude (full square) at negative pump-probe delay t = -130 ps (i.e. t ≅ 13 ns after the previous pump pulse) versus the applied longitudinal magnetic field. The solid red line is a lorentzien fit of the PCD signal. Inset: calculated magnetic field dependence of the average ensemble spin polarization at t = 13 ns, for typical values of $|\beta|=0.6$, $g_h$ =1.5 and $T_\Delta^h = 10 ns$ (see text).

Figure 3 – Time-resolved PL intensity after a $\sigma^+$ laser excitation for co- (I$^+$) and counter-polarized (I$^-$) detection (solid line). We extract a trion lifetime $T_R \approx 800$ ps. The dotted line represents the corresponding circular polarization dynamics, with a characteristic decay time $T_\Delta^e$ ~500 ps. The excitation energy is 1.43 eV and the detection energy, 1.36 eV, is fixed at the centre of the time-integrated PL spectrum (see inset).

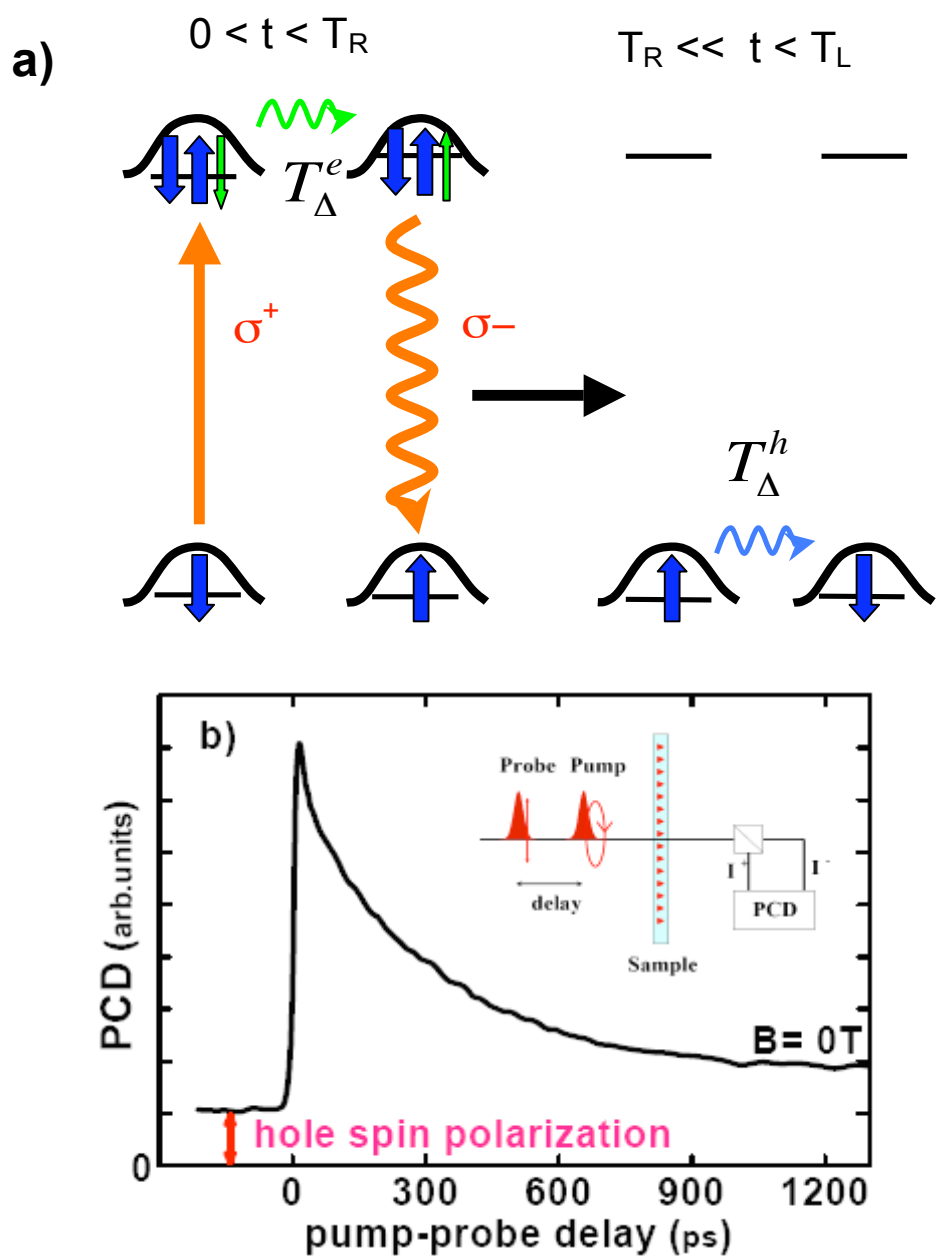

FIG 1

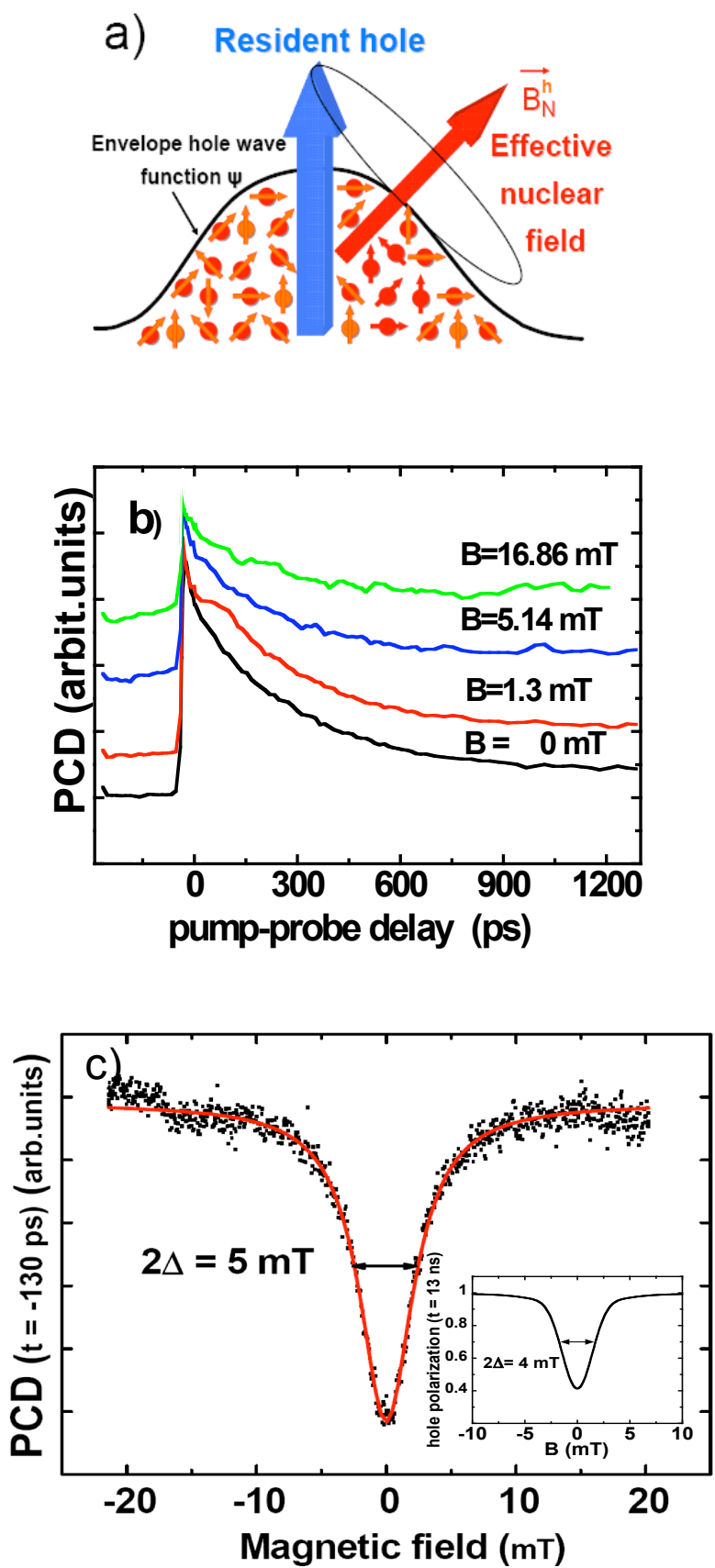

FIG 2

17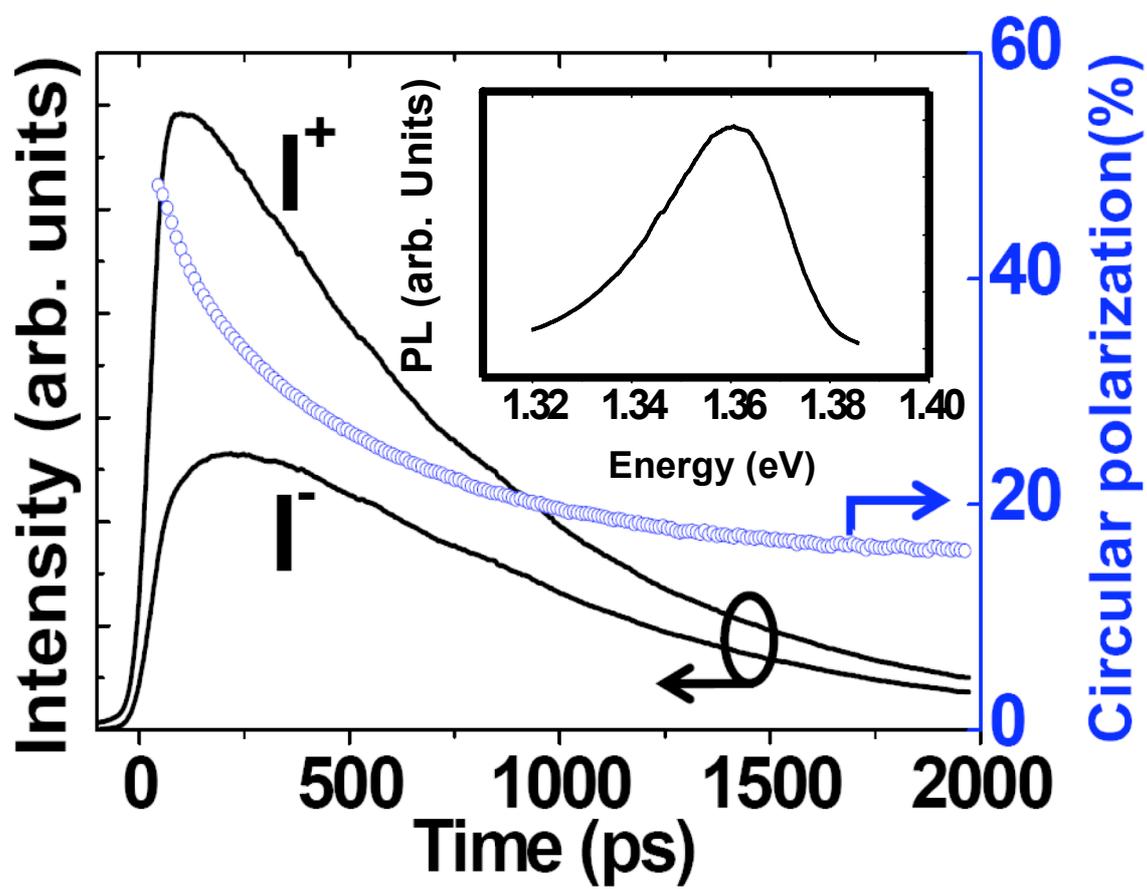

FIG 3